\documentclass[preprintnumbers,amsmath,amssymb,showpacs, twocolumn]{revtex4}

\usepackage{graphicx}
\usepackage{dcolumn}
\usepackage{bm}

\begin{document}

\title{ Teleporting a quantum state in a subset of the whole Hilbert space}

\author{  Mei-Yu Wang $^{1}$, Feng-Li Yan $^{1,2}$}

\affiliation {$^1$ College of Physics Science and Information Engineering, Hebei Normal University, Shijiazhuang 050016, China  \\
$^2$ CCAST (World Laboratory), P.O. Box 8730, Beijing 100080, China}

\date{\today}

\begin{abstract}
We investigate the lower bound of the amount of entanglement for faithfully teleporting a quantum state
belonging to a subset of the whole Hilbert space. Moreover, when the quantum state belongs to a set composed of
two states, a probabilistic teleportation scheme is presented using a non-maximally entangled state as the
quantum channel. We also calculate the average transmission efficiency of this scheme.
\end{abstract}

\pacs{03.67.Hk}

\maketitle

\section{Introduction}

Quantum entanglement is one of the most striking features of quantum mechanics and has been widely used as an
essential resource in the quantum information processing. Some physical incidents such as quantum teleportation
\cite {Bennett93}, quantum key distribution \cite {Ekert}, quantum computation \cite {NielsenChuang,
CiracZoller, Barenco} and quantum secure direct communication \cite {ShimizuImoto, Beige, DengLong, YanZhang,
Gao} are all use it essentially. Since quantum entanglement is the essential resource, we always expect to use
less entanglement when completing a task, or to complete more tasks using certain amount of entanglement. The
two-body entanglement $E$ is defined as the Von Neumann entropy of either of the two subsystems A and B:
\begin{equation}
E(|\Psi\rangle)=-{\rm Tr}(\rho_A{\rm log}_2\rho_A)=-{\rm Tr}(\rho_B{\rm log}_2\rho_B),
\end{equation}
where $ \rho_A$ is the partial trace of $|\Psi\rangle\langle\Psi|$ over subsystem B, and $ \rho_B$ has a similar
meaning. In present paper, we will use it to denote the amount of entanglement.

Since an arbitrary pure state of two-body has the form of Schmidt decomposition:
\begin{equation}
|\Psi\rangle=\sum_{i=1}^{n}\sqrt {p_i}|i\rangle|i\rangle,
\end{equation}
where $n$ is the Schmidt number and $p_i$ is the Schmidt coefficient (we have included in the sum only the
non-zero $p_i$'s). Thus it follows easily that $E=-\sum_{i}p_i{\rm log}_2p_i.$ Particularly, $E={\rm log}_2d$
when $|\Psi\rangle$ is the maximally entangled state, where $d$ is the dimension of Hilbert space.

Quantum teleportation is one of the most important applications of quantum entanglement. In quantum
teleportation process, an unknown quantum state can be transmitted from a sender (called Alice) to a receiver
(Bob) without transmission of carrier of quantum state. Since Bennett et al \cite {Bennett93} presented a
quantum teleportation scheme, there have been great development in theoretical and experimental studies. Now
quantum teleportation has been generalized to many cases such as continuous variable quantum teleportation \cite
{Vaidman}, probabilistic teleportation \cite {LiLiGuo, LuGuo, YanTanYang, YanBai, GaoYanWang,
GaoWangYanChinPhysLett}, controlled quantum teleportation \cite {ZhouHouWuZhang, YanWang,Gao2} and so on \cite
{Bandyopadhyay, Rigolin, ShiGuo, ZengLong, MorHorodecki,LiuGuo, DaiLi, GaoYanWangChinesePhysics}. Moreover,
quantum teleportation has been demonstrated with the polarization photon \cite {BPZ} and a single coherent mode
of fields \cite {Boschi}
 in the experiments. The teleportation  of a coherent state
corresponding to continuous variable system was also  realized in the laboratory \cite {Furusawa}.

In the Bennett's protocol \cite {Bennett93}, the quantum state to be teleported belongs to the whole Hilbert
state vector space. A quantum state of $d$-state particle (or qudit) can be faithfully teleported using a pair
of $d$-state particle in a maximally entangled state, in which, entanglement $E={\rm log}_2 d$  is used. Nielsen
pointed out that it is minimal entanglement for faithfully teleporting an arbitrary $d$ dimensional quantum
state \cite {Nielsen}. When knowing the quantum state in the subspace, we can complete the quantum teleportation
using less entanglement. For instance, Gorbachev and Trubilko \cite {GorbachevTrubilko} considered the quantum
teleportation of two-particle entangled state by a three-particle GHZ state. The entanglement required is
$E={\log}_22=1$ instead of the $\log_24=2$ entanglement for teleporting a general two-qubit state. Yan and Yang
discussed the  economical teleportation of multiparticle quantum state \cite {YanYang}.

In this paper, we will study how many entanglement must be used at least when the quantum state belongs to a
subset of the whole Hilbert space. The lower bound of entanglement for completing faithful teleportation in this
case is calculated. Moreover, when we know the quantum state is coming from a two-state set, a probabilistic
teleportation scheme is presented using a non-maximally entangled state as the quantum channel. The transmission
efficiency of this scheme is calculated also.

\section{The lower bound for teleporting a quantum state in a subset of the whole Hilbert space}

Suppose the quantum state to be teleported belongs to a set $S=\{|\phi_i\rangle, i=1,2,\cdots\}$, which is a
subset of the $d$ dimensional Hilbert space either finite or infinite. In the following, we will investigate the
lower bound of entanglement when teleporting a quantum state from $S$.

{\bf Case 1}: The set $S$ is an orthogonal set, i.e. the arbitrary two quantum states in $S=\{|\phi_i\rangle,
i=1,2,\cdots\}$ are orthogonal.

Apparently, $S$ must be the finite set in this case. Let the number of quantum states be $n$. Alice can know
exactly what state she has by measuring the quantum state to be teleported in the orthogonal basis
$\{|\phi_1\rangle,|\phi_2\rangle,\cdots, |\phi_n\rangle\}$. Then she simply sends Bob classical information
saying which state it is, and Bob may prepare it himself, i.e. in this process Alice and Bob do not need  any
entanglement for teleporting the state, thus $E=0$.

{\bf Case 2}: The set $S$ is a non-orthogonal set, i.e. at least a pair states in the set  $S=\{|\phi_i\rangle,
i=1,2,\cdots\}$ are non-orthogonal.

Evidently, we can search for a maximum linear independent subset of $S$ denoted by
$S'=\{|\phi_{i_1}\rangle,|\phi_{i_2}\rangle,\cdots,|\phi_{i_m}\rangle\}$, other states in $S$ may be expressed
as the linear combination of $|\phi_{i_1}\rangle,|\phi_{i_2}\rangle,\cdots,|\phi_{i_m}\rangle$. Suppose that we
can teleport an arbitrary quantum state in $S$ via the quantum channel $|\psi\rangle_{23}$. So the state in
$\{|\phi_{i_1}\rangle,|\phi_{i_2}\rangle,\cdots,|\phi_{i_m}\rangle\}$ can be teleported of course.

In the following we will prove inverse conclusion: If one can teleport an arbitrary quantum state in $S'$ via
the quantum channel $|\psi\rangle_{23}$, then one can teleport an arbitrary quantum state in $S$.

Considering  a general process of teleportation, let quantum channel $|\psi\rangle_{23}$ can teleport quantum
state $|\phi\rangle_1$. Then the state of the whole system composed of the state of the particle to be
teleported and quantum channel is written as
\begin{equation}
|\Psi\rangle_{123}=|\phi\rangle_1|\psi\rangle_{23}.
\end{equation}
By the hypothesis that  quantum channel $|\psi\rangle_{23}$ can teleport quantum state $|\phi\rangle$, there
must exist a decomposition,
\begin{equation}
|\Psi\rangle_{123}=|\phi\rangle_1|\psi\rangle_{23}=\sum_k|k\rangle_{12}U_k^{-1}|\phi\rangle_3,
\end{equation}
where $ \{|k\rangle, k=1,2,\cdots, r\}$ is an orthogonal basis of particles 1 and 2. When Alice performs a joint
projective measurement on particles 1 and 2 in the basis $ \{|k\rangle, k=1,2,\cdots, r\}$, particle 3 will
collapse into the state $U_k^{-1}|\phi\rangle_3$. Then Alice sends the measurement outcome to Bob via a
classical channel. After receiving  Alice's message, Bob performs a unitary operation $U_k$ on his particle 3
according to Alice's measurement outcome. The quantum state of particle 3 will be transformed into
$|\phi\rangle$ which Alice wants to teleport, thus teleportation is achieved. Note that Alice does not know what
state to be teleportated  before teleportation, so the unitary transformation $U_k^{-1}$ and $U_k$ must be
independent of $|\phi\rangle$. Since the every state in
$\{|\phi_{i_1}\rangle,|\phi_{i_2}\rangle,\cdots,|\phi_{i_m}\rangle\}$ can be teleported using the above quantum
channel, so
\begin{equation}
\begin{array}{l}
|\phi_{i_1}\rangle_1|\psi\rangle_{23}=\sum_k|k\rangle_{12}U_k^{-1}|\phi_{i_1}\rangle_3,\\
|\phi_{i_2}\rangle_1|\psi\rangle_{23}=\sum_k|k\rangle_{12}U_k^{-1}|\phi_{i_2}\rangle_3,\\
\cdots \cdots\cdots\\
|\phi_{i_m}\rangle_1|\psi\rangle_{23}=\sum_k|k\rangle_{12}U_k^{-1}|\phi_{i_m}\rangle_3.
\end{array}
\end{equation}
Therefore,
\begin{equation}
\sum_{i_m}c_{i_m}|\phi_{i_m}\rangle_1|\psi\rangle_{23}=\sum_k|k\rangle_{12}U_k^{-1}\sum_{i_m}c_{i_m}|\phi_{i_m}\rangle_3,
\end{equation}
where $c_{i_m}$ is an arbitrary complex. Eq.(6) shows that the arbitrary linear combination of
$\{|\phi_{i_1}\rangle,|\phi_{i_2}\rangle,\cdots,|\phi_{i_m}\rangle\}$ can be teleported by the entangled channel
$|\psi\rangle_{23}$, in other words, we may teleport an arbitrary quantum state in the linear space by states in
$S$ and vice versa.

That is, the entanglement is the same when we teleport either an arbitrary quantum state in $S$ or a one in
$m$-dimensional linear space by the states in $S$. According to Nielsen's theorem, the minimum entanglement in
this case is $E=\log_2 m$, where $m$ is the number of quantum state of maximum linear independent subset of $S$.
Evidently, $m\leq d$, the amount of entanglement is less than that for teleporting a quantum state from the
whole Hilbert space.

From the above discussions, we have drawn two conclusions:

1. If the quantum states in $S$ is orthogonal, we need not any entanglement for teleportation, but as long as we
plus a state into $S$, which is at least non-orthogonal with one in $S$, then the cost of entanglement becomes
$\log_2 m$ suddenly.

2. In spite of knowing more knowledge about the quantum state to be teleported, for example, we know it is one
of the two states, if the two states are non-orthogonal, we can not complete the teleportation with less than
the full unit of entanglement.

\section{A Probabilistic Teleportation Scheme}

In the above section, we show that if we know the state belongs to a subset of the whole Hilbert space, we can
complete the faithful teleportation with less entanglement than that of teleporting a quantum state of the whole
Hilbert space.

In the other hand, for given entanglement,  is it possible to teleport more quantum states when we know the
state is in a subset of the whole Hilbert space? When the quantum channel is a non-maximally entangled state,
the answer is positive. In the following, we will present a probabilistic teleportation scheme for the case
where the quantum state to be teleported is from a two-state set. The transmission efficiency of this  scheme is
calculated.

Let us denote the quantum channel by
\begin{equation}
|\psi\rangle_{23}=x|00\rangle_{23}+y|11\rangle_{23},
\end{equation}
where $x,y$ are real and satisfy $|x|<|y|$. The quantum state to be teleported is in
$\{|\phi_1\rangle,|\phi_2\rangle\}$ $(\langle\phi_1|\phi_2\rangle=A{\rm e}^{i\theta})$. In order to realize the
teleportation, Alice introduces an auxiliary qubit $a$ with the original state $|0\rangle_a$. So the initial
state of particle 1,2,3 and $a$ is
\begin{equation}
|\Psi\rangle_{1a23}=|\psi\rangle_1|0\rangle_a(x|00\rangle_{23}+y|11\rangle_{23}),
\end{equation}
where $|\psi\rangle_1 \in \{|\phi_1\rangle,|\phi_2\rangle\}$. Alice performs a unitary transformation
\begin{equation}
U_{a2}=\left [\begin{array}{cccc} 1& 0& 0 & 0\\
0& x/y & 0 &\sqrt {1-x^2/y^2}\\
0 & 0 &1& 0\\
0&\sqrt {1-x^2/y^2}& 0 &-x/y
\end{array}\right],
\end{equation}
on particles $a$ and 2. Correspondingly $|\Psi\rangle_{1a23}$ becomes
\begin{eqnarray}
&& ~~~I_1\otimes U_{a2}\otimes I_3|\Psi\rangle_{1a23}\nonumber\\
&&=\sqrt 2 x|\psi\rangle_1|0\rangle_a\frac {1}{\sqrt 2}{(|00\rangle_{23}+|11\rangle_{23})}\nonumber\\
&&~~~+\sqrt {1-2x^2}|\psi\rangle_1|1\rangle_a|11\rangle_{23}.
\end{eqnarray}
Then Alice measures the auxiliary particle $a$. If the result is $|0\rangle_a$, the state of particle 2 and 3 is
the maximally entangled state, thus, Alice can teleport $|\psi\rangle_1$ successfully by the standard
teleportation procedure, and the success probability is $2x^2$. When the result $|1\rangle_a$ occurs, Alice
makes a POVM containing three elements,
\begin{eqnarray}
&&E_1\equiv \frac {1}{1+|\langle\phi_1|\phi_2\rangle|}|\phi_1^\perp\rangle\langle\phi_1^\perp|,\\
&&E_2\equiv \frac {1}{1+|\langle\phi_1|\phi_2\rangle|}|\phi_2^\perp\rangle\langle\phi_2^\perp|,\\
&&E_3\equiv I-E_1-E_2,
\end{eqnarray}
on the unknown quantum state. Here $|\phi_1^\perp\rangle$ and $|\phi_2^\perp\rangle$ are chosen by the relation
$\langle\phi_1|\phi_1^\perp\rangle=0$ and $\langle\phi_2|\phi_2^\perp\rangle=0$ respectively. It is
straightforward to verify that the above operators are positive operators which satisfy the completeness
relation $\sum_m E_m=I$, and therefore form a legitimate POVM.

After Alice performs the measurement described by the POVM $\{E_1, E_2, E_3\}$, if the result of her measurement
is $E_1$, then Alice can safely conclude that the state must be $|\phi_2\rangle$, because there is zero
probability that the state is $|\phi_1\rangle$, since $E_1$ has been cleverly chosen to ensure that
$\langle\phi_1|E_1|\phi_1\rangle=0$. A similar line of reasoning shows that if the measurement outcome $E_2$
occurs then the state must be $\phi_1$. However, some of the time, Alice will obtain the measurement outcomes
$E_3$, and Alice can infer nothing about the identity of the state. The key point, however, is that Alice never
makes a mistake identifying the state. This infallibility comes at the price that sometimes Alice obtains no
information about the identity of the state. Evidently, if the measurement outcome $E_1$ (or $E_2$) occurs,
Alice simply sends Bob classical information saying which state it is, and Bob may prepare it himself.  If the
measurement result is $E_3$, the teleportation fails.

It is easy to calculate that the probability to successfully distinguish the quantum state $|\phi_1\rangle$ and
$|\phi_2\rangle$ is
\begin{equation}
p=1-|\langle\phi_1|\phi_2\rangle|.
\end{equation}

 Synthesizing all   cases, the probability of successful
teleportation in this scheme is
\begin{equation}
p_{suc}=2x^2+(1-|\langle\phi_1|\phi_2\rangle|)(1-2x^2).
\end{equation}
i.e. the average number of the quantum states to be teleported via each partially entangled channel is
\begin{equation}
I_{tran}=2x^2+(1-|\langle\phi_1|\phi_2\rangle|)(1-2x^2).
\end{equation}

When $x=1/\sqrt 2,$ the partially entangled channel becomes the maximally one, whatever
$|\langle\phi_1|\phi_2\rangle|$ it is, $p_{suc}=1$. If $\langle\phi_1|\phi_2\rangle=0$, the two states are
orthogonal, whatever $x$ it is, $p_{suc}=1$. These two cases correspond  two special limitations.

\section{Conclusion}

In conclusion, we have found the lower bound of the amount of quantum entanglement required for faithfully
teleporting  a quantum state in a subset of the whole Hilbert space. Moreover, when the quantum state belongs to
a two-state set, a probabilistic teleportation scheme is presented using a non-maximally entangled state as the
quantum channel. The average transmission efficiency of this scheme is obtained also.

\acknowledgments This work was supported by Hebei Natural Science Foundation of China under Grant Nos:
A2004000141 and A2005000140, and  Natural Science Foundation of Hebei Normal University.

\end{document}